\documentclass[brazil,english,aps,prl,twocolumn,superscriptaddress,showpacs,noshowkeys]{revtex4}
\usepackage{lmodern}
\usepackage[T1]{fontenc}
\usepackage[latin9]{inputenc}
\usepackage{color}
\usepackage{babel}
\usepackage{units}
\usepackage{graphicx}
\usepackage{xcolor}
\usepackage{soul}
\usepackage[unicode=true,
 bookmarks=false,
 breaklinks=false,pdfborder={0 0 1},backref=false,colorlinks=false]
 {hyperref}
\hypersetup{
 colorlinks,citecolor=blue,filecolor=black,linkcolor=red,urlcolor=blue}
\usepackage{breakurl}
\usepackage{epstopdf}
\usepackage{makecell}

\makeatletter

\@ifundefined{textcolor}{}
{%
 \definecolor{BLACK}{gray}{0}
 \definecolor{WHITE}{gray}{1}
 \definecolor{RED}{rgb}{1,0,0}
 \definecolor{GREEN}{rgb}{0,1,0}
 \definecolor{BLUE}{rgb}{0,0,1}
 \definecolor{CYAN}{cmyk}{1,0,0,0}
 \definecolor{MAGENTA}{cmyk}{0,1,0,0}
 \definecolor{YELLOW}{cmyk}{0,0,1,0}
}

\usepackage[babel=true]{microtype}
\usepackage{amsfonts}
\usepackage{braket}
\usepackage{tensor}
\usepackage{amsmath}
\usepackage{amssymb}
\usepackage{gensymb}
\usepackage{bbold}

\makeatother

\begin{document}

\title{Optical simulation of the free Dirac Equation}

\author{Thais L. Silva}

\email{thaisdelimasilva@gmail.com}

\affiliation{Instituto de Física, Universidade Federal do Rio de Janeiro, 21941-972,
Rio de Janeiro, Rio de Janeiro, Brazil}

\author{E. R. F. Taillebois}

\affiliation{Instituto de Física, Universidade Federal de Goiás, 74.690-900, Goiânia,
Goiás, Brazil}

\author{R. M. Gomes}

\affiliation{Instituto de Física, Universidade Federal de Goiás, 74.690-900, Goiânia,
Goiás, Brazil}

\author{S. P. Walborn}

\affiliation{Instituto de Física, Universidade Federal do Rio de Janeiro, 21941-972,
Rio de Janeiro, Rio de Janeiro, Brazil}

\author{Ardiley T. Avelar}

\affiliation{Instituto de Física, Universidade Federal de Goiás, 74.690-900, Goiânia,
Goiás, Brazil}

\begin{abstract}
We present a classical optics simulation of the one-dimensional Dirac equation for a free particle.   Positive and negative energy components are represented by orthogonal polarizations of a free propagating beam, while the spatial profile represents the spatial wave function of the particle.  Using a completely tunable time parameter, we observed the oscillation of the average value of the Dirac position operator--known as \emph{Zitterbewegung} (ZB).  We are also able to measure the so called mean-position operator which is a single-particle observable and presents no oscillations.  Our work opens the way for optical simulation  of interesting phenomenon of relativistic systems, as well as condensed-matter physics, without any requirement for specially engineered medium.
\end{abstract}
\maketitle

\section{Introduction}

\noindent \qquad{}Although the Dirac equation \cite{Dirac1928} represents a historical landmark in the quantum description of relativistic systems -- satisfactorily explaining the origin of spin and predicting the existence of antimatter \cite{Anderson1933} -- 
it still provokes a lot of discussion about its interpretation, even when applied to describe the simplest physical system, that is, a free particle. In this case, the Dirac equation predicts intriguing phenomena, for instance the ZB \cite{Schrodinger1930} and Klein's paradox \cite{Klein1929}, which impede the single-particle (SP) interpretation of the Dirac equation.  As fundamental effects in the understanding of relativistic influence over quantum theory, they have contributed to the transition to the many-body quantum field approach \cite{Krekora04}.

The important technical difficulties involved in the direct observation of several relativistic quantum predictions have led to an increased interest in their simulations in trapped ions \cite{Gerritsma10,Gerritsma11}, photonic crystals \cite{Zhang08}, confined light \cite{Otterbach09}, graphene \cite{Katsnelson06}, optical superlattices \cite{Dreisow10}, Bose-Einstein condensates \cite{Salger11,LeBlanc13} and ultracold atoms \cite{Vaishnav08}. Among the unexpected effects of Dirac equation, the ZB -- the flickering motion of a free relativistic quantum particle described by a Dirac wavefunction with positive and negative energy components -- is one of most investigated due to its interesting counterintuitive nature.

Despite the previous studies \cite{Gerritsma10,Dreisow10,LeBlanc13}, an important feature was not explored: the Dirac position operator related to the flickering motion is not a SP observable, i.e. it cannot be written as the direct sum of its positive and negative energy projections. From a SP perspective, Dirac's equation must be interpreted as the simultaneous solving of two independent problems, for instance, the single free evolution of both electron and positron. Therefore, there is no meaning in assigning a physical interpretation to an operator that is not SP. This fact can be well understood by making explicit the connection between the theory of irreducible unitary representations of the Poincaré group and the explicitly covariant reducible representation associated to Dirac's equation \cite{Caban03}. In the SP approach, physical results must be obtained by projecting SP observables over the subspace corresponding to the problem of interest.

For Dirac's theory, a SP position observable exists and is obtained using the so-called Foldy--Wouthuysen transformation (FWT) \cite{Foldy50} -- a momentum dependent unitary transformation that diagonalizes Dirac's Hamiltonian and is at the kernel of important algorithms used to obtain quantum relativistic corrections \cite{Obukhov01,Quach15}. This observable is often called mean-position operator and does not exhibit the oscillatory behavior characteristic of the ZB, a result that generates doubts concerning its actual existence.

Here, the simulation of the Dirac one-dimensional free evolution and the ZB is performed using the transverse degrees of freedom of a paraxial light beam, where different components of the spinor are represented by different polarization components of the beam. This physical setup is well suited for the purpose of quantum simulation, as it allows for implementation of the dynamical phases with easy tuning of the important physical parameters \cite{Lemos12}. Besides being a proof-of-concept for the optical simulation of Dirac particles, the present approach differs from others in the theoretical procedure adopted to perform the simulation. Previous works perform a direct simulation of the 1+1D Dirac Hamiltonian, while the present approach performs the evolution in the diagonalized Foldy--Wouthuysen representation (FWR) and permits one to switch back and forth between this and Dirac's representation, allowing us to investigate the behavior of both the Dirac position and the mean-position operators.

\section{Dirac Equation and position operator}

Consider the 1D Dirac equation
\begin{equation} 
i\hbar\frac{\partial}{\partial t}\psi_{D}=\hat{\mathcal{H}}_{D}\psi_{D}=(c\hat{p}\sigma_{1}+mc^{2}\sigma_{3})\psi_{D},\label{dirac}
\end{equation}
where $c$ is the speed of light, $\hat{p}$ is the momentum operator, $m$ is the mass of the particle, and $\sigma_{i}$ are the usual Pauli matrices. The information of this system is encoded in the spinor $\psi_{D}$ that has only two components which are related to positive and negative energy states in the particle's rest frame, i.e. spin degrees of freedom are eliminated by the dimensional constrain \cite{Schwabl}. In Dirac's coordinate representation, the momentum operator $\hat{p}$ assumes the usual form $-i\hbar\frac{\partial}{\partial x}$, where $x$ is the so-called Dirac coordinate associated to the multiplication operator $\hat{x}_{D}\psi_{D}(x)=x\psi_{D}(x)$. Since the Hamiltonian operator is not diagonal in this representation, the positive and negative energy eigenstates are non-trivial and assume, respectively, the forms $\psi_{p}^{+}(x,t)=u(p)e^{-i\varepsilon(p)t/\hbar}e^{ipx/\hbar}$ and $\psi_{p}^{-}(x,t)=v(p)e^{i\varepsilon(p)t/\hbar}e^{-ipx/\hbar}$, with $\varepsilon(p)\equiv\sqrt{(mc^{2})^{2}+(pc)^{2}}$, $u(p)=[2mc^{2}(\varepsilon(p)+mc^{2})]^{-1/2}\begin{pmatrix}\varepsilon(p)+mc^{2} & cp\end{pmatrix}^{T}$ and $v(p)=[2mc^{2}(\varepsilon(p)+mc^{2})]^{-1/2}\begin{pmatrix}cp & \varepsilon(p)+mc^{2}\end{pmatrix}^{T}$.

The non-diagonal form of $\hat{\mathcal{H}}_{D}$ in Dirac's representation is evinced by the commutator $[\hat{\mathcal{H}}_{D},\hat{x}_{D}]=-ic\hbar\sigma_{1}$ and leads to the Heisenberg picture evolution given by \cite{Thaller}
\begin{equation}
\begin{aligned}
\hat{x}_{D}(t)= & \hat{x}_{D}(0)+c^{2}\hat{p}\hat{\mathcal{H}}_{D}^{-1}t \\
 & -\frac{c\hbar\hat{\mathcal{H}}_{D}^{-1}}{2i}\left(e^{2i\hat{\mathcal{H}}_{D}t/\hbar}-1\right)\left(c\hat{p}\hat{\mathcal{H}}_{D}^{-1}-\sigma_{1}\right).
\end{aligned}
\label{eqEvol}
\end{equation}
The first two terms on the right represent the expected linear time evolution of a free particle, the last term being associated to the ZB. This flickering motion is accompanied by other particularities of the $\hat{x}_{D}$ operator. Indeed, the evolution given in \eqref{eqEvol} is derived from the equation of motion $\dot{\hat{x}}_{D}=\frac{i}{\hbar}[\hat{\mathcal{H}}_{D},\hat{x}_{D}]=c\sigma_{1},$ which implies that, although $\langle\dot{\hat{x}}_{D}\rangle=\langle c^{2}\hat{p}\hat{\mathcal{H}}_{D}^{-1}\rangle$, the eigenvalues associated to the velocity ${\dot{\hat{x}}_{D}}$ are restricted to $\pm c$, a remarkable result which contributes to raise doubts as to the correct interpretation of $\hat{x}_{D}$ as definition of position. These peculiarities of the operator $\hat{x}_{D}$ arise from the fact that this is not a SP observable, i.e. $\hat{x}_{D}\neq\hat{P}_{+}\hat{x}_{D}\hat{P}_{+}^{\dagger}+\hat{P}_{-}\hat{x}_{D}\hat{P}_{-}^{\dagger}$, where $\hat{P}_{\epsilon}=\frac{1}{2mc^{2}}\left(\begin{smallmatrix}mc^{2}+\epsilon\varepsilon(p) & -\epsilon cp\\ \epsilon cp & mc^{2}-\epsilon\varepsilon(p) \end{smallmatrix}\right)$ is the projection operator over the subspace of states with energy sign $\epsilon$.

To obtain a SP position, the FWT must be applied to diagonalize the Dirac Hamiltonian. For the 1D Dirac free particle, this canonical transformation is given by the momentum dependent unitary operator $\hat{U}(\hat{p})=e^{i\hat{S}(\hat{p})}$ with $\hat{S}(\hat{p})\equiv\frac{\sigma_{2}}{2}\mathrm{tg}^{-1}\left(\frac{\hat{p}}{mc}\right)$. In the resulting FWR, the original Dirac Hamiltonian is given by $\hat{\mathcal{H}}_{D}^{\prime}=\sigma_{3}\varepsilon(\hat{p})$, and the former $\hat{x}_{D}$ operator by $\hat{x}_{D}^{\prime}=\hat{x}_{FW}+\frac{\hbar mc^{3}}{2\varepsilon(p)^{2}}\sigma_{2}$, where $\hat{x}_{FW}$ is the new multiplication operator in the FWR. The operator $\hat{x}_{FW}$ is the so called mean-position operator and, unlike the operator $\hat{x}_{D}$, it is a SP observable since $\hat{x}_{FW}=\hat{P}_{+}^{\prime}\hat{x}_{FW}\hat{P}_{+}^{\prime}+\hat{P}_{+}^{\prime}\hat{x}_{FW}\hat{P}_{+}^{\prime}$, where $\hat{P}_{\epsilon}^{\prime}=\left(\begin{smallmatrix}\delta_{\epsilon+} & 0\\ 0 & \delta_{\epsilon-} \end{smallmatrix}\right)$ are the energy projectors in the new representation.

Aside from being SP, the operator $\hat{x}_{FW}$ also satisfies the equation $\dot{\hat{x}}_{FW}=c^{2}\hat{p}\hat{\mathcal{H}}_{D}^{\prime}$, resulting in the Heisenberg picture evolution \cite{Thaller}
\begin{equation}
\hat{x}_{FW}(t)=\hat{x}_{FW}(0)+c^{2}\hat{p}\hat{\mathcal{H}}_{D}^{\prime  -1}t
\end{equation}
that is linear in time, as expected for a free particle. Thus, as stated before, the ZB does not occur for this operator.

Here, as a proof-of-concept for the simulation of relativistic systems using free propagating light beams, the simulation of both the Dirac and FWRs will be performed in a single setup. This difference with other simulation procedures open the possibility for future investigations on more complex FWTs associated to relativistic scenarios involving interactions.

\section{Simulation Protocol and Experiment}

One way to simulate the dynamics associated to Eq. \eqref{dirac} is to directly implement the evolution operator $\exp\left(-\frac{i\hat{\mathcal{H}}_{D}t}{\hbar}\right)$, which is usually a tough task due to the non-diagonal character of $\hat{\mathcal{H}}_{D}$. This difficulty can be overcome by using the FWT, since this transformation allows the time evolution operator to be written as $\exp\left(-\frac{i\hat{\mathcal{H}}_{D}t}{\hbar}\right)=\hat{U}^{-1}(\hat{p})\exp\left(-\frac{i\hat{\mathcal{H}}_{D}^{\prime}t}{\hbar}\right)\hat{U}(\hat{p}).\label{ev_op}$ This operator can be implemented in an optical beam by considering the vertical coordinate on the transverse plane as the particle's position and the horizontal (vertical) polarization as the superior (inferior) component of the spinor. The horizontal spatial degrees of freedom on the transverse plane play no relevant role in the experiment. Although a spinor is a mathematical object which transforms very specifically under a reference frame change, it is not a concern for this simulation since the reference frame is assumed to be fixed.

The optical transformations required for the simulation are polarization transformations (acting as nondiagonal operators) and phase shifts (used to introduce momentum dependent phases). The former are obtained with the suitable application of wave plates and the last are realized by spatial light modulators (SLMs), which are able to imprint programmable position dependent phases $\textrm{exp}[i\:\textrm{f}(x,y)]$ in the horizontal polarization. To transform these position-dependent phases into momentum-dependent ones, optical Fourier transforms are used. The action of a quarter wave plate (QWP) set to $45^{\circ}$ is given by the operator $\hat{Q}=\nicefrac{e^{i\tfrac{\pi}{4}}}{\sqrt{2}}\left(\mathbb{1}-i\sigma_{1}\right)$, while $\hat{H}=\sigma_{1}$ describes a half wave plate at $45^{\circ}$. The action of a SLM is equivalent to applying $\hat{P}[f(x,y)]=\textrm{exp}[i\:\textrm{f}(x,y)]\sigma_{+}\sigma_{-}+\sigma_{-}\sigma_{+}$ over the transverse profile spinor. Using this operator representation for the optical devices, it follows that 
\begin{equation}
\hat{U}(p)=\hat{Q}\hat{P}\left[-\theta(p)\right]\hat{H}\hat{P}\left[\theta(p)\right]\hat{Q},
\end{equation}
with $2\theta(p)=\mathrm{tan}^{-1}\left(\frac{p}{mc}\right)$. We express the inverse FWT in an analogous fashion. As the Hamiltonian is diagonal in the FWR, the transformed time evolution operator is achieved via the application of the dynamical phase $\exp\left[\pm it\varepsilon(p)/\hbar\right]$ in each spinor component using waveplates and the SLM, which concludes the simulation. A summary of the analogy between the optical simulator and the simulated system is given in Table I.

\begin{table}
\caption{Summary of the optical analogy}
\begin{tabular}{c|c}
\hline 
\textbf{Optical System} & \textbf{Simulated System}\tabularnewline
\hline 
\hline 
Vertical transverse position & $x$\tabularnewline
\hline 
\makecell{Transverse profile of \\horizontal polarization} & $\psi_{D_{1}}(x)$\tabularnewline
\hline 
\makecell{Transverse profile of \\ vertical polarization} & $\psi_{D_{2}}(x)$\tabularnewline
\hline 
QWP@45\textdegree{} & $\ensuremath{\hat{Q}=\nicefrac{e^{i\tfrac{\pi}{4}}}{\sqrt{2}}\left(\mathbb{1}-i\sigma_{1}\right)}$\tabularnewline
\hline 
HWP@45\textdegree{} & $\hat{H}=\sigma_{1}$\tabularnewline
\hline 
SLM printing phase $f(x,y)$ & \makecell{$\hat{P}[f(x,y)]=$ \\$\textrm{exp}[i\:\textrm{f}(x,y)]\sigma_{+}\sigma_{-}+\sigma_{-}\sigma_{+}$}\tabularnewline
\hline 
\makecell{Normalized horizontal \\ polarization intensity at $x$} & 
$\left|\psi_{D_{1}}(x)\right|^{2}$\tabularnewline
\hline 
\noalign{\vskip0.2cm}
\end{tabular}
\end{table}

\begin{figure}
\includegraphics[width=0.9\columnwidth]{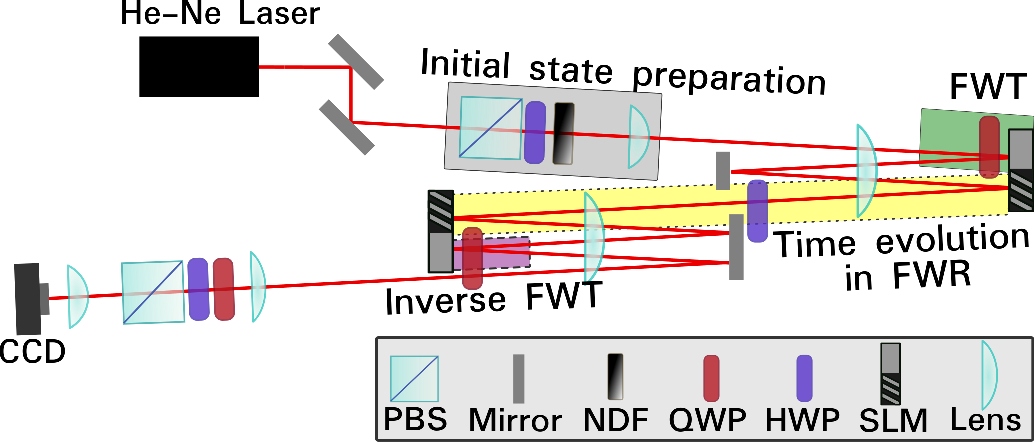}\caption{\label{fig:Experimental-setup}A He-Ne laser, prepared with an initial gaussian profile and anti-diagonal polarization state, is sent through an optical system designed to implement the Dirac Hamiltionian.  The grey shaded regions of the SLMs implement the FWTs, while the striped regions implement the free-evolution.  Lenses are used to map the field profile among the different planes of the SLMs, and wave plates to control the polarization state.  A CCD camera is used to register the intensity profile of the output field.  Additional details are provided in the text.  }
\end{figure}

The experimental scheme is shown in Fig.\ref{fig:Experimental-setup}. A He-Ne laser with wavelength 632.8 nm and two Holoeye reflective SLMs, each of which divided into halves to operate twice, are used. The Fourier transforms are made by plano-convex cylindrical lenses with  150 mm focal distance such that the position space (mirrors and camera)  is in one focal plane and the momentum space is in the opposite focal plane where the SLM is located. The momentum $p$ and the position on the SLM, $X$, are connected by $X=\frac{\lambda f}{h}p$, where $f$ is the focal distance and $\lambda$ is the laser wavelength \cite{Saleh}. In terms of $X$, the applied phases become $2\theta(X)=\mathrm{tan}^{-1}\left(\frac{h}{mc}\frac{X}{\lambda f}\right)$ and $t\varepsilon(p)/\hbar= 2\pi t\sqrt{\left(\frac{X}{\lambda f}\right)^{2}+\left(\frac{mc}{h}\right)^{2}}$, so the parameters we need to set are the speed of light $c$ and the Compton wavelength $\lambda_{C}=h/mc$, which are easily tunable since they enter as programmable parameters in the imprinted phases. Since the time coordinate also comes up as a programmable parameter, we could in
principle take measurements for as many time values as we wish inside a time interval. This also implies that the unit of measurement for time is an arbitrary $\tau$. In this realization we chose $\Delta t/\tau = 1$.

The laser produces a Gaussian spatial profile separable in the $x$ and $y$ coordinates, so the initial spinor is 
\begin{equation}
\psi_{D}(x,t=0)=\frac{e^{-i\pi x^{2}/(\lambda R)}e^{-x^{2}/(4\Delta^{2})}}{(\sqrt{2\pi}2\Delta)^{1/2}}\begin{pmatrix}a\\
b
\end{pmatrix},\label{eq:initial_state}
\end{equation}
where $a$ and $b$ are the proportions of horizontal and vertical polarizations ($|a|^{2}+|b|^{2}=1$), $\Delta$ is the beam width in the vertical direction, and $R$ is the vertical radius of curvature of the beam in the initial position plane. The propagation and Gouy terms of the Gaussian beam only introduce global phases which do not affect the dynamical evolution \cite{Saleh}. We start with $a=-b=1$, but changing $a$ and $b$ would enable us to prepare different positive and negative energy superpositions. Two cylindrical lenses are placed before the first position space in order to manipulate the initial momentum distribution which depends on $R$ and therefore on $\Delta$. Using a beam profiler, we determined the initial state parameters to be $\Delta=48,6\,\mu m $ and $\lambda R/\pi=(2.2\Delta)^2$.

The average position of the simulated particle is calculated as 
$
\langle\hat{x}_{D}\rangle(t)=\sum_{i=1,2}\int dx\,x|\psi_{Di}(x,t)|^{2},$ 
where $\left|\psi_{Di}(x,t)\right|^{2}$ is proportional to the light intensity of polarization component $i$ at position $x$ on the transverse plane measured by a CCD camera placed at the output position space. Each instant of time corresponds to one programmable-phases configuration and one intensity-profile measurement. It is worth noting that the evolved state is accessible for any time value.

\section{Results}
\begin{figure}
\includegraphics[width=0.95\columnwidth]{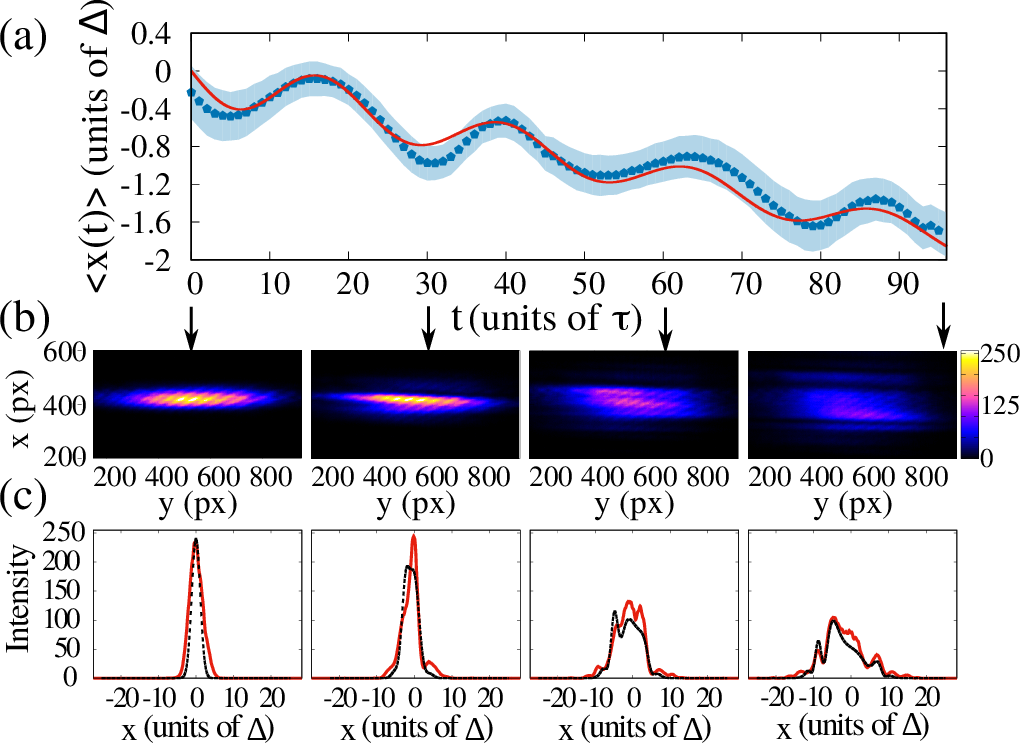}\caption{\label{fig:L5Results-1} (a) Mean position $\langle\hat{x}_{D}(t)\rangle$ as a function of the time parameter $t$ for Compton length $\lambda_{C}/\Delta=5$. Points are experimental data, the dashed line is the numerical solution and filled band is the experimental error of one $\sigma$. (b) The camera image  and (c) the one dimensional state got from marginalizing the intensity measured by the camera over $y$ for four values of the parameter $t$, namely $t/\tau=0,\:30,\:60\textrm{ and }95$, are shown. In (c) red solid lines are experimental data and black dashed curves are numerical solutions. Intensity is in grayscale.}
\end{figure}

A summary of our experimental procedure for the particular case $\lambda_{C}=5\Delta$ is depicted in Fig.\ref{fig:L5Results-1}. In Fig.\ref{fig:L5Results-1}-(a) we present the mean position $\langle\hat{x}_{D}\rangle(t)$ as a function of $t$, the ZB being evidenced by the oscillatory behavior. The solid red line is the theoretical prediction, while points are experimental results obtained from the images shown in Fig.\ref{fig:L5Results-1}-(b).  The shaded region represents uncertainty of one $\sigma$. Fig.\ref{fig:L5Results-1}-(b) shows samples of the data collected by the CCD camera for some instants of time, the $x$ distribution being obtained by considering only a fixed $y$ coordinate at the center of the beam. The $x$ distributions used to calculate $\langle\hat{x}_{D}\rangle(t)$ are shown in Fig.\ref{fig:L5Results-1}-(c). This procedure assumes that $x$ and $y$ intensity distributions remain separable troughout all the apparatus. This is true in the ideal case, however the cylindrical lenses can introduce some non-separability as one can see in the slightly tilted eliptical intensity pattern shown in Fig. \ref{fig:L5Results-1}-(b). The non-separability causes the initial state to be not entirely pure. Since our experimental results agree well with theory, we conclude that these effects are negligible for the present experiment.

\begin{figure}
\begin{center}
 \includegraphics[width=0.9\columnwidth]{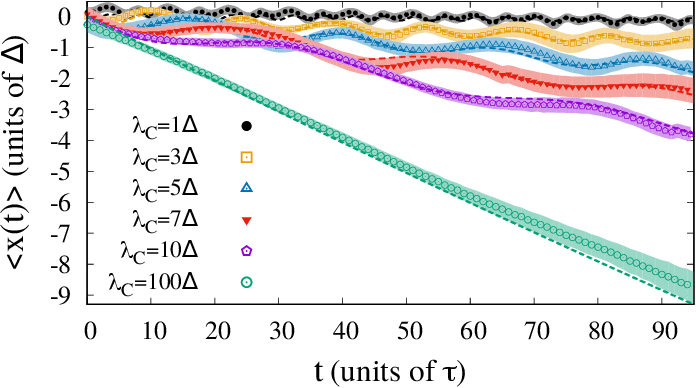}
\end{center}
\vspace{-0.5cm}
\caption{\label{fig:AllMeanPosition} Average position $\langle\hat{x}_{D}(t)\rangle$ as a function of the time parameter $t$ for Compton lengths $\lambda_{c}/\Delta=1,3,5,7,10,100$. Dashed lines are numerical predictions of the theory using Eq. \eqref{dirac}, while points are experimental results. The filled bands are experimental errors of one $\sigma$.}
\end{figure}

\begin{figure}
\includegraphics[width=0.9\columnwidth]{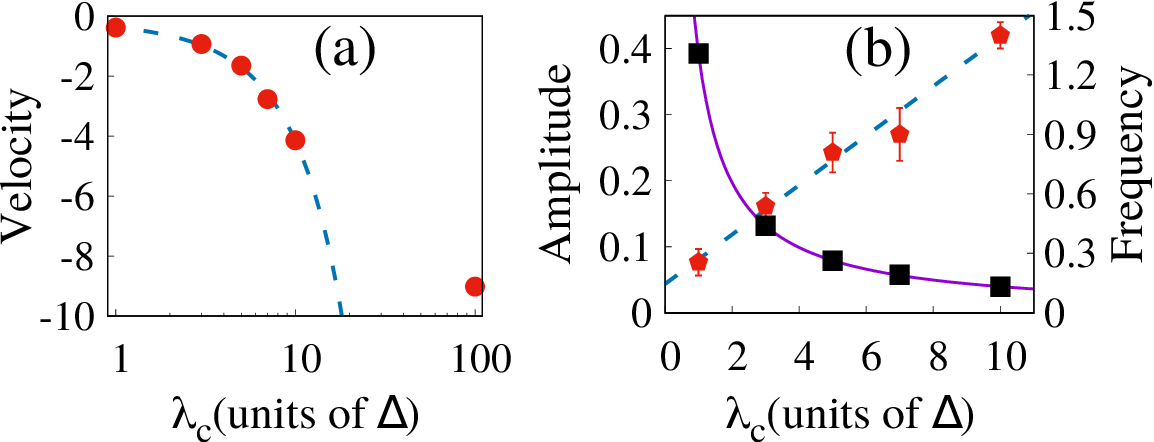}
\caption{\label{fig:amp_freq} (a) Mean velocity (in units of $10^{-2}\Delta/\tau$), (b) amplitude (red circles, in units of $\Delta$) and frequency (black squares, in units of $\tau^{-1}$) of ZB obtained by fitting a sinusoidal function to the average position $\langle\hat{x}_{D}(t)\rangle$ for each Compton wavelength.  For small values of Compton wavelength (large masses) the mean velocity falls quadratically with $\lambda_{C}$ (dashed blue line) and for a large value of Compton wavelength (vanishing mass) it appoximates the speed of light set on the experiment. The amplitude dependence with $\lambda_{C}$ is in well agreement with linear behavior (dashed blue line) while frequency is proportional to $1/\lambda_{c}$ (solid purple line). }
\end{figure}

For a fixed initial state and speed of light $c=0.1\,\nicefrac{\Delta}{\tau}$, we measured the average position in Dirac's representation for different values of the Compton wavelength, as shown in Fig.\ref{fig:AllMeanPosition}. We fitted the average position with the function $vt+A\sin(\omega t+\delta)$ for each $\lambda_{C}$ to estimate the mean velocity, amplitude and frequency of the oscillation. These quantities are shown in Fig.\ref{fig:amp_freq}. Our experimental results are in agreement with the expected linear dependence of amplitude and inverse dependence of frequency on $\lambda_{C}$ for small $\lambda_{C}$ \cite{nota}, as can be seen in Fig.\ref{fig:amp_freq}-(b). This is consistent with the fact that the ZB visibility in Fig.\ref{fig:AllMeanPosition} increases for smaller values of $\lambda_C$.

The different inclinations exhibited in Fig.\ref{fig:AllMeanPosition} are due to the fact that each mass, i.e. Compton wavelength, is associated to a different velocity distribution, even the momentum distribution being the same for all values of $\lambda_{C}$. Although the initial state \eqref{eq:initial_state} has zero average
momentum, this is not true for the mean velocity in Fig.\ref{fig:amp_freq}-(a). As is expected, the mean velocity falls quadratically with $\lambda_{C}$ for large masses (small $\lambda_{C}$), while it is close to the speed of light for very small masses ($\lambda_{C}=100\Delta$). 

The agreement between the experimental ZB data and the theoretical predictions confirm that our optical setup is well suited for the study of 1+1D relativistic dynamical systems, the theoretical extension to larger dimensions being discussed in the Appendix. Beside serving as a proof-of-concept, the proposed setup permits to investigate the system in the FWR, an interesting possibility since it allows to describe the dynamics of the system according to the single-particle perspective, i.e. assigning physical sense only to the projections of single-particle operators over the subspaces of definite sign of energy.

From the SP perspective, operators that are not block-diagonal in the FWR, as is the case for $\hat{x}_{D}$, have no physical meaning, since they mix components of positive and negative energy that are associated to two distinct problems. On the other hand, operators that are block-diagonal in the FWR, as $\hat{x}_{FW}$ or the Dirac Hamiltonian, may have a physical sense assigned to their positive and negative projections. In this sense, the correct description of the dynamics of a single electron (positron), for example, should be given by $\hat{x}_{FW,+} = P^{\prime}_{+}\hat{x}_{FW}P^{\prime}_{+}$ ($\hat{x}_{FW,-} = P^{\prime}_{-}\hat{x}_{FW}P^{\prime}_{-}$). In our setup, the average $\langle\hat{x}_{FW}\rangle$ measured using both positive and negative components of the spinor can be obtained by measuring the transverse profile of the beam before the inverse FWT. However, since the positive and negative components of the spinor are encoded in the horizontal and vertical polarizations of the beam in the FWR, the single-particle position dynamics described by $\hat{x}_{FW,+}$ (particle) and $\hat{x}_{FW,-}$ (anti-particle) is also accessible by simply selecting one of the polarizations prior to the CCD measurement in the FWR.

\begin{figure}[!h]
\includegraphics[width=0.8\columnwidth]{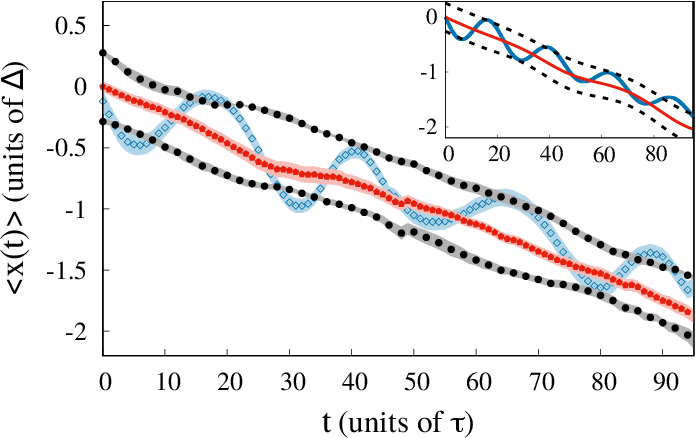}\caption{\label{fig:L5FWMeanPosition} Experimental mean positions $\langle\hat{x}_{D}(t)\rangle$ (blue dots) and $\langle\hat{x}_{FW}(t)\rangle$ (red circles) for $\lambda_{C}/\Delta=5$. Measurements of $\hat{x}_{FW}(t)$ projected over positive energy components (horizontal polarization in FWR) as well as negative energy components (vertical polarization in FWR) are shown as black dots. The shaded regions are the mean value uncertainties.  The $\langle\hat{x}_{FW}(t)\rangle$ data presents no ZB and fits a linear dependence with $R^2=99.6\%$. The inset picture shows the result of a numeric simulation of the experimental setup assuming ideal devices except that the SLMs have efficiency of $95\%$. The inset axes are the same as the main picture.}
\end{figure}

Experimental results for the mean-position operator are shown in Fig. \ref{fig:L5FWMeanPosition} for $\lambda_C=5 \Delta$. The experimental data concerning the ZB effect for the $\hat{x}_{D}$ operator is also plotted for comparison (blue points). Measurements of  $\langle\hat{x}_{FW,+}(t)\rangle$ and $\langle\hat{x}_{FW,-}(t)\rangle$ are shown as black dots. As is expected from the independence of the two problems in the SP description, we have two independent mean trajectories corresponding to the free evolution of the particle and the corresponding anti-particle. The ZB is not present for these mean trajectories and a linear behavior in time is observed, as it was expected. As mentioned earlier, we were also able to measure the mean value of the FW mean-position operator $\hat{x}_{FW}$. The results are plotted as the red circles and represent an average of the positive and negative projections cases. The small deviation from a perfect linear behavior can be explained assuming that the SLMs do not modulate all the incident light but a fraction of it, as is shown in the inset picture which shows the same mean values as the experimental plot but obtained from a numerical simulation of the experiment for modulation efficiency of $95\%$ in each SLM.

\section{Discussion and Conclusions}

Our experiment demonstrates how relativistic dynamics can be studied using classical optics, and opens the way to more sophisticated investigations. For this purpose it would also be desirable to produce more general initial states. This can be accomplished using intensity and phase masks in the initial state preparation. The state produced in this experiment had zero avarage momentum, but simply shifting the momentum in all SLMs phases by the same $\Delta p$ can be interpreted as if the state has non-vanishing average momentum.

In principle the method implemented in this simulation using the FWT could be applied for other simulation schemes of the Dirac equation, however this transformation requires applying a phase shift that is proportional to the inverse tangent of momentum. In our approach the application of this phase is fairly easy, thanks to the spatial light modulator (SLM). However, in other systems, this is quite challenging.  Typically in continuous variable quantum simulators one can implement Gaussian Hamiltonians, but non-Gaussian operations (third order and above) are quite difficult \cite{Tasca11}. Thus, we believe that our approach is quite interesting in this regard, as it allows one to employ the FWT and investigate relevant aspects of it.

Albeit here we present the theory and results for the 1+1D case, the extension for 2+1D and for some initial states in 3+1D is straightforward as shown in the Appendix. The first is a direct extension considering the second transverse coordinate of the beam as the second spatial degree of freedom of the simulated particle. A 2+1D simulation also allows for investigation of electronic behavior in bidimensional condensed matter systems such as graphene \cite{Katsnelson06}, but still do not present any spin effect. For a general 3+1D simulation it would be necessary a third beam coordinate what is not available in this scheme. In spite of this limitation, in the Appendix we show a class of initial states which dependence on the third coordinate does not alter the time evolution. In this case, the two extra spinor components are provided by the polarization components of a second beam. 

It is well known that there is no exact FWT for the non-free Dirac equation, i.e., if we add a potential to the free Dirac equation \eqref{dirac} the Dirac Hamiltonian becomes no longer diagonalizable with one single unitary transformation \cite{Schwabl}. This seems to be a very limiting factor of our simulation technique and indeed it is if we try to implement the actual FWT for a potential problem. Instead of doing so, we can try to find other kinds of unitary transformations which reproduce the time evolution and are experimentaly feasible with the available optical elements. Up to now we know that at least for a particular class of potentials it is possible to break the time evolution operator into a position dependent phase which carries all the information about the potential followed by the free evolution presented in this work. This particular case is discussed on Appendix B.

In conclusion, we have presented an all-optical simulation of the dynamics of a one-dimensional relativistic free point particle, where the beam's spatial profile plays the role of the particle's wavefunction, and its orthogonal polarization components are associated to spinor components.  Our experiment is based on the diagonalization of the Dirac Hamiltonian using the FWT, which allowed for the decomposition of the unitary evolution into operations that are realizable with off-the-shelf optical components. Adjusting the tunable time parameter we observed the oscillatory ZB phenomenon for Dirac's position operator. Using our experimental FWT, we were also able to address this phenomenon from a single-particle perspective, where the position description is given by the positive and negative energy projections of the single-particle mean-position operator. This approach allowed us to observe the absence of ZB oscillations for the particle and anti-particle single-particle dynamical evolutions.

\begin{acknowledgments}
We thank the Brazilian agencies CAPES (PROCAD2013 project), CNPq (\#459339/2014-1), FAPEG (PRONEX \# 201710267000503) and the Instituto Nacional de Ciência e Tecnologia-Informação Quântica (INCT-IQ) for partial
support. 
\end{acknowledgments}

\appendix

\section{ Simulation for more spatial degrees of freedom}

The goal of this Appendix is to show the simulation protocols for particles in two and three spatial dimensions.

\subsection{Simulation of 2+1 dimensional Dirac Equation}

Consider the Dirac Hamiltonian for a free particle existing in a 2D space
\begin{equation}
H=c(\sigma_{1}p_{x}+\sigma_{2}\mathrm{p}_{y})+mc^{2}\sigma_{3}.
\end{equation}
The unitary transformation $e^{iS_2}=e^{-i\frac{\sigma_{1}p_{x}-\sigma_{2}p_{y}}{|\mathbf{p}|}\theta(\mathbf{p})}$
is the FWT which diagonalizes the Hamiltonian in this case, with $\theta(\mathbf{p})=\frac{1}{2}\textrm{tg}^{-1}\frac{|\mathbf{p}|}{mc}$.
For the same reasons as in the 1D situation, there is no spin if the
space is restricted to two dimensions and the particle state is a
spinor with two components. Once we manage to construct FWT from optical
device operators, the simulation protocol is made possible identifying
again the transverse profile of a laser beam in the two orthogonal
polarizations with the spinor components and identifying the two transverse
coordinates with particle position. To show that in fact there is
such a decomposition let us define a new momentum dependent phase
\begin{equation}
\theta'(\mathbf{p})=\begin{cases}
\mathrm{tg}^{-1}(p_{y}/p_{x}), & p_{y}>0\textrm{ or }p_{y}=0,p_{x}>0\\
\pi+\mathrm{tg}^{-1}(p_{y}/p_{x}), & p_{y}<0\textrm{ or }p_{y}=0,p_{x}<0,
\end{cases}\label{eq:thetalinha}
\end{equation}
motivated by the polar expression $p_{x}-ip_{y}=|\mathbf{p}|e^{-i\theta'}$.
In terms of the two phase functions the FWT reads
\begin{equation}
e^{iS_2}=\left(\begin{array}{cc}
\cos\theta(\mathbf{p}) & e^{-i\theta'}\sin\theta(\mathbf{p})\\
-e^{i\theta'}\sin\theta(\mathbf{p}) & \cos\theta(\mathbf{p})
\end{array}\right).\label{eq:TFW2D}
\end{equation}
Using the definitions given in the main text it is easy to see that
the operator sequence $\hat{P}\left[-\theta^{'}(\mathbf{p})\right]\hat{Q}\hat{P}\left[-\theta(\mathbf{p})\right]\hat{H}\hat{P}\left[\theta(\mathbf{p})\right]\hat{Q}\hat{P}\left[\theta^{'}(\mathbf{p})\right]$
is equal to the FW unitary. The transformed diagonalized time evolution
is then given by the product $\hat{H}\hat{P}\left[i\varepsilon(\mathbf{p})t/\hbar\right]\hat{H}\hat{P}\left[-i\varepsilon(\mathbf{p})t/\hbar\right]$.
Just to conclude the protocol, the inverse FWT is given analogously
by the same set of devices as the FWT with different imprinted phases
and different angles for the wave plates.

\subsection{Simulation of 3+1 dimensional Dirac Equation for a particular class of initial states}

Let us consider the 3+1 Dirac equation
\begin{equation}
 i\hbar \frac{\partial \psi(\mathbf{x},t)}{\partial t}=\left[c\boldsymbol{\alpha}\cdot \boldsymbol{p}+mc^2\beta\right]\psi(\mathbf{x},t)
\end{equation}
with the standard choice of Dirac matrices $\beta=\begin{pmatrix}\mathbb{1}_{2\times2} & 0\\
0 & -\mathbb{1}_{2\times2}
\end{pmatrix}\quad,\qquad\alpha_{k}=\begin{pmatrix}0 & \sigma_{k}\\
\sigma_{k} & 0
\end{pmatrix}$. The four components of the spinor $\psi(\mathbf{x},t)$ accounts for the two signs of the energy and for the two spin projections along a fixed direction. The FWT reads $e^{iS_3}=\cos{\theta(\mathbf{p})}+\beta\frac{\boldsymbol{\alpha\cdot}\mathbf{p}}{|\mathbf{p}|}\mathrm{sen}\,\theta(\mathbf{p})$ with the same definition for $\theta(\mathbf{p})$ as before.

A general simulation of the above equation, besides of requiring a four dimensional object to emulate the four spinor components, it  would also require three spatial degrees of freedom, while the presented setup allows for just two. Instead of proposing a complete simulation, let us consider only the particular family of initial states given by
\begin{equation}
\psi(\mathbf{x},t=0)=\begin{cases}
\frac{1}{\sqrt{L}}\left(\begin{array}{c}
\tilde{\phi}_{1}(x,y)\\
\vdots\\
\tilde{\phi}_{4}(x,y)
\end{array}\right) & -\frac{L}{2}<z<\frac{L}{2}\\
0 & \mathrm{elsewhere}
\end{cases}.
\end{equation}
In momentum space this state becomes
\begin{equation}
\tilde{\psi}(t=0,\boldsymbol{p})=\left(\begin{array}{c}
\tilde{\phi}_{1}(p_{x},p_{y})\\
\vdots\\
\tilde{\phi}_{4}(p_{x},p_{y})
\end{array}\right)\sqrt{\frac{L}{2\pi\hbar}}\left(\frac{\sin{\frac{p_{z}L}{2\hbar}}}{\frac{p_{z}L}{2\hbar}}\right),\end{equation}
which depedence on $p_z$ behaves like a $\delta(p_z)$ for large values of $L$.

Thus we can approximate the FW transformed state by
\begin{widetext}
\begin{equation}
 \tilde{\psi}'(\boldsymbol{p},t=0)\approx\left(\cos{\left(\theta(p_x,p_y,0)\right)}+\beta\frac{\alpha_{x}p_{x}+\alpha_{y}p_{y}}{|(p_x,p_y,0)|}\sin{\theta(p_x,p_y,0)}\right)\tilde{\psi}(\boldsymbol{p},t=0),
\end{equation}
\end{widetext}
and all the momentum dependent phases only depend on two coordinates and can be aplied with SLMs.

The transformed state is explicitly written as
\begin{widetext}
\begin{equation}
 \psi'(\boldsymbol{p},t=0)=\left(\begin{array}{cccc}
\cos{\theta} & 0 & 0 & \frac{p_{x}-ip_{y}}{|\mathbf{p}|}\sin{\theta}\\
0 & \cos{\theta} & \frac{p_{x}+ip_{y}}{|\mathbf{p}|}\sin{\theta} & 0\\
0 & \frac{-p_{x}+ip_{y}}{|\mathbf{p}|}\sin{\theta} & \cos{\theta} & 0\\
\frac{-p_{x}-ip_{y}}{|\mathbf{p}|}\sin{\theta} & 0 & 0 & \cos{\theta}
\end{array}\right)\left(\begin{array}{c}
\phi_{1}\\
\phi_{2}\\
\phi_{3}\\
\phi_{4}
\end{array}\right)\sqrt{\frac{L}{2\pi\hbar}}\left(\frac{\sin{\frac{p_{z}L}{2\hbar}}}{\frac{p_{z}L}{2\hbar}}\right)
\end{equation}
\end{widetext}
we notice that the FWR only mixes the components two by two what makes possible to simulate it using two beams without any joint transformation between them. Then we can address the transverse profiles of horizontal and vertical polarizations of the first beam to $\phi_1$ and $\phi_4$ and the FWR as well as the subsequent diagonal time evolution do not mix this components with the two remaining. Moreover, each pair of mixed components transforms like the two spatial dimensions case (Eq. \eqref{eq:TFW2D}) with the suitable phase signs.

The interesting thing about three spatial dimensions simulation is that it would enable us to investigate also spin effects like the spin analogous to \emph{Zitterbewegung}.

\section{Simulation of 1+1 dimensional Dirac Equation for a particular class of potentials}

Employing the strategy of  \cite{Sabin12},  we show in this Appendix that, if the initial state is conveniently prepared, our approach is able to simulate the Dirac equation for a particular class of potentials.  To this end,  consider the 1D Dirac equation
\begin{equation}
i\hbar\frac{\partial}{\partial t}\psi_{D}=(c\hat{p}\sigma_{1}+mc^{2}\sigma_{3}+V(x))\psi_{D},\label{dirac2}
\end{equation}
where $V(x)$ is a spinorial potential of the form
\begin{equation}
V(x) = V_1(x)\sigma_1.
\end{equation}

For the above particular case, we define $\phi_D$ such that
\begin{equation}
\psi_D=e^{-i\frac{1}{\hbar c}\int V_1(x^{\prime}) \mathbb{1}dx^{\prime}}\phi_D.
\end{equation}
The substitution of this state on Eq.(\ref{dirac2}) shows that the spinor $\phi_D$ evolves according to the free Dirac equation (Eq.(\ref{dirac})). Thus, if we want to simulate the time evolution of the initial state $\psi_D(x,t=0)$, we need to prepare the state $\phi_D(x,t=0)=\exp \left[i\frac{1}{\hbar c}\int V_1(x^{\prime}) \mathbb{1}dx^{\prime}\right] \psi_D(x,t=0)$ and the dynamics of the free evolution.
From the experimental point of view, this corresponds to applying a local phase in position space to all the components of the spinor before performing the free evolution in the way as it is shown in the main text.


\begin{thebibliography}{1}
\bibitem{nota} From Eq.\eqref{eqEvol} we have that the amplitude of ocillations for a given momentum eigenstate is proportional to $ch/\epsilon(p)$, which for small $\lambda_C$, i.e., large mass, can be approximated by $\lambda_C$. On the other hand, the frequency is proportional to $\epsilon(p)/h$, that approaches $c/\lambda_C$ for large masses.
\end{thebibliography}
\end{document}